\documentclass[journal,draft,onecolumn,twoside]{IEEEtran}
\normalsize

\newcommand{\av}{\ensuremath{\mathbf{a}}}

\newcommand{\hv}{\ensuremath{\mathbf{h}}}
\newcommand{\mv}{\ensuremath{\mathbf{m}}}

\newcommand{\pv}{\ensuremath{\mathbf{p}}}

\newcommand{\vv}{\ensuremath{\mathbf{v}}}

\newcommand{\wv}{\ensuremath{\mathbf{w}}}
\newcommand{\xv}{\ensuremath{\mathbf{x}}}
\newcommand{\yv}{\ensuremath{\mathbf{y}}}
\newcommand{\zv}{\ensuremath{\mathbf{z}}}

\newcommand{\Am}{\ensuremath{\mathbf{A}}}

\newcommand{\Ym}{\ensuremath{\mathbf{Y}}}
\newcommand{\Gammam}{\ensuremath{\mathbf{\Gamma}}}
\newcommand{\Hm}{\ensuremath{\mathbf{H}}}
\newcommand{\Zm}{\ensuremath{\mathbf{Z}}}

\def\Pr{\mathrm{Pr}}

\DeclareMathAlphabet{\mcl}{OMS}{cmsy}{m}{n}

\newlength\tikzwidth
\newlength\tikzheight

\usepackage{amsmath}
\usepackage{amssymb}
\usepackage{tikz}
\usetikzlibrary{arrows,shapes}
\usepackage{pgfplots}
\pgfplotsset{compat=1.3}
\usepgflibrary{shapes}
\usepackage{algorithm}
\usepackage[noend]{algpseudocode}
\usepackage{bbm}
\makeatletter
\def\BState{\State\hskip-\ALG@thistlm}
\makeatother

\definecolor{mycolor1}{rgb}{0.63529,0.07843,0.18431}%
\definecolor{mycolor2}{rgb}{0.00000,0.44706,0.74118}%
\definecolor{mycolor3}{rgb}{0.00000,0.49804,0.00000}%
\definecolor{mycolor4}{rgb}{0.87059,0.49020,0.00000}%
\definecolor{mycolor5}{rgb}{0.00000,0.44700,0.74100}%
\definecolor{mycolor6}{rgb}{0.74902,0.00000,0.74902}%

\begin{document}

\title{An Enhanced Decoding Algorithm for Coded Compressed Sensing with Applications to Unsourced Random Access}
\author{\IEEEauthorblockN{
    Vamsi K. Amalladinne,
    Jamison R. Ebert,
    Jean-Francois Chamberland, and
    Krishna R. Narayanan
    }
    \thanks{
    This material is based on work supported, in part, by the National Science Foundation (NSF) under Grants CCF-1619085 \& CCF-2131106, and by Qualcomm Technologies, Inc., through their University Relations Program.
    }
}

\maketitle
\begin{abstract}
Unsourced random access (URA) has emerged as a pragmatic framework for next-generation distributed sensor networks.
Within URA, concatenated coding structures are often employed to ensure that the central base station can accurately recover the set of sent codewords during a given transmission period. 
Many URA algorithms employ independent inner and outer decoders, which can help reduce computational complexity at the expense of a decay in performance.
In this article, an enhanced decoding algorithm is presented for a concatenated coding structure consisting of a wide range of inner codes and an outer tree-based code. It is shown that this algorithmic enhancement has the potential to simultaneously improve error performance and decrease the computational complexity of the decoder.
This enhanced decoding algorithm is applied to two existing URA algorithms and the performance benefits of the algorithm are characterized.
Findings are supported by numerical simulations. 
\end{abstract}


\section{Introduction}
\label{section:Introduction}

Massive machine-type communication (mMTC) is a rapidly growing class of wireless communications which aims to connect tens of billions of unattended devices to wireless networks. 
One significant application of mMTC is that of distributed sensing, which consists of a large number of wireless sensors that gather data over time and transmit their data to a central server, which then interprets the received data to produce useful information and/or make executive decisions. 
When combined with recent advances in machine learning (ML), such networks are expected to open a vast realm of economic and academic opportunities.
However, the large population of unattended devices within these networks threatens to overwhelm existing wireless communication infrastructures by dramatically increasing the number of network connections; it is expected that the number of machines connected to wireless networks will exceed the population of the planet by an entire order of magnitude.
Additionally, the traffic and demand profiles characteristic of individual sensors and actuators are highly inefficient under existing human-centric communication protocols; specifically, the sporadic and bursty nature of sensor transmissions are very costly under estimation/enrollment/scheduling procedures typical of cellular networks. 
The combination of these challenges necessitates the design of novel physical and medium access control (MAC) layer protocols to efficiently handle the demands of these wireless devices. 

One recently proposed paradigm for efficiently handling the demands of unattended devices is that of unsourced random access (URA), first proposed by Polyanskiy in 2017 \cite{polyanskiy2017perspective}. 
URA captures many of the nuances of IoT devices by considering a network with an exceedingly large number of uncoordinated devices, of which, only a small percentage is active at any given point in time. 
When a device/user is active, it encodes its short message using a common codebook and then transmits its codeword over a regularly scheduled time slot, as facilitated by a beacon. 
Furthermore, the power available to each user is strictly limited and assumed to be uniform across devices. 
The use of a common codebook is characteristic of URA and has two important implications: first, the network does not need to maintain a dictionary of active devices and their unique codebook information; second, the receiver does not know which node transmitted a given message unless the message itself contains a unique identifier. 
The receiver is then tasked with recovering an unordered list of transmitted messages sent during each time slot by the collection of active devices. 
The performance of URA schemes is evaluated with respect to the per-user probability of error (PUPE), which is the probability that a user's message is not present in the receiver's final list of decoded messages (this measure is defined in \eqref{eq:pupe}).
In \cite{polyanskiy2017perspective}, Polyanskiy provides finite block length achievability bounds for the short block lengths typical of URA applications using random Gaussian coding and maximum likelihood (ML) decoding. 
However, these bounds were produced in the absence of complexity constraints and thus are impractical for deployment in real-world networks. 
Over the past few years, several URA schemes have been proposed as means to obtain near-optimal performance with tractable complexity \cite{ordentlich2017isit, vem2017user, pradhan2020sparse, marshakov2019polar, pradhan2019polar, pradhan2019joint, pradhan2021ldpc, amalladinne2019coded, fengler2019sparcs, amalladinne2020unsourced, calderbank2018chirrup, ebert2020hybrid, ebert2021stochastic, decurninge2020tensorbased, shyianov2020massive, han2021sparsekron,li2020sparcldpc, xie2020correlatedmimo, liang2021iterativemimo, fengler2019massive, fengler2019mimo, truhachev2021lowcomplexura}.

All of the aforementioned URA schemes employ concatenated channel codes to recover the messages sent by the collection of active users at the receiver. 
We note that the term \emph{channel code} is used broadly such that it includes certain signal dictionaries such as those commonly used for compressed sensing (CS). 
Though it is conceptually simpler to decode the inner and outer codes independently, it is a well-known fact within coding theory that dynamically sharing information between the inner and outer decoders will often improve the performance of the decoder. 
In this paper, we present a novel algorithm for sharing information between a wide class of inner codes and a tree-based outer code that significantly improves the PUPE performance and reduces the computational complexity of the scheme.
Specifically, our main contributions are as follows.
\begin{enumerate}
    \item A general system model consisting of a wide class of inner codes and an outer tree code is developed. 
    An enhanced decoding algorithm is presented whereby the outer tree code may guide the convergence of the inner code by restricting the search space of the inner decoder to parity consistent paths.
    \item The coded compressed sensing (CCS) scheme of Amalladinne et al. in \cite{amalladinne2019coded} is considered under this model. 
    The enhanced decoding algorithm is applied to CCS and the performance benefits are quantified. 
    \item The CCS for massive MIMO scheme of Fengler et al. in \cite{fengler2019mimo} is considered under this model. 
    The enhanced decoding algorithm is applied to CCS for massive MIMO and the performance benefits are quantified. 
\end{enumerate}


\section{System Model}
\label{section:SystemModel}

Consider a URA system consisting of $K$ active devices which are referred to by a fixed but arbitrary label $j \in [K]$. 
Each of these users wishes to simultaneously transmit their $B$ bit message $\wv_j$ to a central base station over a Gaussian multiple access channel (GMAC) using a concatenated code consisting of an inner code $\mathcal{C}$ and an outer tree code $\mathcal{T}$.
This inner code $\mathcal{C}$ has the crucial property that, given a linear combination of $K \leq \delta$ codewords, the constituent information messages may be individually recovered with high probability. 
Furthermore, we assume that the probability that any two active users' messages are identical is low, i.e. $\Pr(\wv_i = \wv_j) < \epsilon$ for $i \neq j$.

We consider a scenario where it is either computationally intractable to inner encode/decode the entire message simultaneously or it is otherwise impractical to transmit the entire inner codeword at once; thus, each user must divide its information message into fragments and inner encode/decode each fragment individually.
To ensure that the message can be reconstructed from its fragments at the receiver, the information fragments are first connected together using an outer tree-based code $\mathcal{T}$, and then inner-encoded using code $\mathcal{C}$.
The resulting signal is transmitted over the channel.
We elaborate on this process below.

\begin{figure}[htb]
    \centering
    \begin{tikzpicture}[
  font=\small, line width=0.75pt,
  infobits0/.style={rectangle, minimum height=7mm, minimum width=25mm, draw=black, fill=gray!10},
  infobits/.style={rectangle, minimum height=7mm, minimum width=15mm, draw=black, fill=gray!10},
  paritybits/.style={rectangle, minimum height=7mm, minimum width=10mm, draw=black, fill=lightgray!50}
]

\node[infobits0] (vb0) at (1.25,0) {$\wv(1)$};
\node[infobits] (vb1) at (3.25,0) {$\wv(2)$};
\node[paritybits] (vp1) at (4.5,0) {$\pv(2)$};
\node[infobits] (vb2) at (5.75,0) {$\wv(3)$};
\node[paritybits] (vp2) at (7.0,0) {$\pv(3)$};
\node[auto] (vdots) at (8.5, 0) {$\hdots$};
\node[infobits] (vb3) at (10.25,0) {$\wv(L)$};
\node[paritybits] (vp3) at (11.5,0) {$\pv(L)$};
\draw[|-|] (0,-0.6) to node[midway,below] {$m_1$} (2.5,-0.6);
\draw[-|] (2.5,-0.6) to node[midway,below] {$m_2$} (4.0,-0.6);
\draw[-|] (4.0,-0.6) to node[midway,below] {$l_2$} (5,-0.6);
\draw[-|] (5,-0.6) to node[midway,below] {$m_3$} (6.5,-0.6);
\draw[-|] (6.5,-0.6) to node[midway,below] {$l_3$} (7.5,-0.6);
\draw[|-|] (9.5,-0.6) to node[midway,below] {$m_{L}$} (11.0,-0.6);
\draw[-|] (11.0,-0.6) to node[midway,below] {$l_{L}$} (12.0,-0.6);
\end{tikzpicture}
    \caption{This figure illustrates the structure of a user's outer encoded message, denoted by $\vv$. Fragment $\ell$ consists of the concatenation of information bits, denoted by $\wv(\ell)$, and parity bits, denoted by $\pv(\ell)$.}
    \label{fig:info_parity_subblocks}
\end{figure}
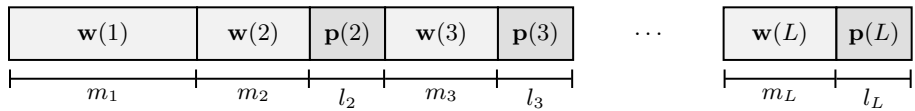

Each message $\wv_j$ is broken into $L$ fragments where fragment $\ell$ has length $m_{\ell}$ and $\sum_{\ell \in [L]} m_{\ell} = B$.
Notationally, $\wv_j$ is represented as the concatenation of fragments by $\wv_j = \wv_j(1)\wv_j(2)\hdots\wv_j(L)$. 
The fragments are outer-encoded together by adding parity bits to the end of each fragment, with the exception of the first fragment.
This is accomplished by taking random linear combinations of the information bits contained in previous sections. 
The parity bits appended to the end of section $\ell$ are denoted by $\pv_j(\ell)$, and it has length $l_{\ell}$.
This outer-encoded vector is denoted by $\vv_j$, where $\vv_j(\ell) = \wv_j(\ell)\pv_j(\ell)$. 
The vector $\vv_j$ now assumes the form shown in Fig.~\ref{fig:info_parity_subblocks}. 

After the outer-encoding process is complete, user~$j$ inner-encodes each fragment $\vv_j(\ell)$ individually using $\mathcal{C}$ and concatenates the encoded fragments to form signal $\xv_{j}$. 
Each user then simultaneously transmits its signal to the base station over a GMAC.
The received signal at the base station assumes the form 
\begin{equation}
    \yv = \sum_{j \in [K]} d \xv_j + \zv
\end{equation}
where $\zv$ is a vector of Gaussian noise with independent standard normal components and $d$ accounts for the transmit power. 

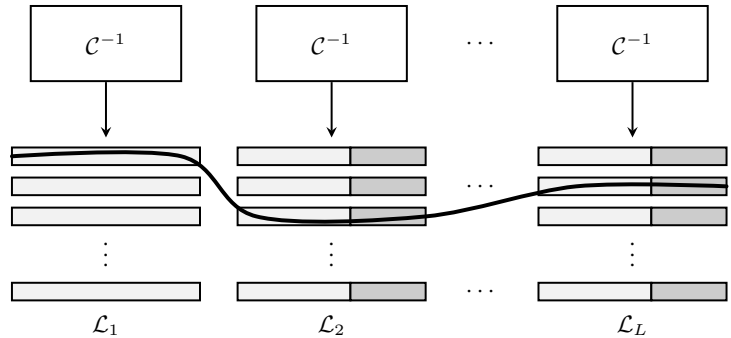
\begin{figure}[htb]
    \centering
    \begin{tikzpicture}[
  font=\small, >=stealth', line width = 0.75pt,
  wblock/.style={rectangle, minimum height=2mm, draw=black, fill=gray!10},
  pblock/.style={rectangle, minimum height=2mm, draw=black, fill=gray!40},
  decoder/.style={rectangle, minimum height=10mm, draw=black, minimum width=20mm}
]




\node[decoder] (dec0) at (1.25, 0.5) {$\mathcal{C}^{-1}$};
\node[decoder] (dec1) at (4.25, 0.5) {$\mathcal{C}^{-1}$};
\node[decoder] (dec2) at (8.25, 0.5) {$\mathcal{C}^{-1}$};
\node at (6.25, 0.5) {$\cdots$};

\draw[-stealth] (dec0) -- (1.25, -0.75);
\draw[-stealth] (dec1) -- (4.25, -0.75);
\draw[-stealth] (dec2) -- (8.25, -0.75);

\foreach \w in {-1.0,-1.4,-1.8, -2.8} {
  \node[wblock,minimum width=25mm] (w0-\w) at (1.25,\w) {};
  \node[wblock,minimum width=15mm] (w1-\w) at (3.75,\w) {};
  \node[pblock,minimum width=10mm] (p1-\w) at (5.0,\w) {};
  \node[wblock,minimum width=15mm] (wn-\w) at (7.75,\w) {};
  \node[pblock,minimum width=10mm] (pn-\w) at (9.0,\w) {};
}
\node at (6.25, -1.4) {$\cdots$};
\node at (6.25, -2.8) {$\cdots$};

\foreach \w in {1.25, 4.25, 8.25} {
  \node at (\w, -2.2) {$\vdots$};
}

\node (L1) at (1.25,-3.25) {$\mathcal{L}_1$};
\node (L2) at (4.25,-3.25) {$\mathcal{L}_2$};
\node (Ln) at (8.25,-3.25) {$\mathcal{L}_{L}$};

\draw[line width=1.5pt,color=black,line cap=round] plot[smooth, tension=.55] coordinates {(0,-1.0) (2.25,-1) (3.25,-1.8) (5.5, -1.8) (7.5,-1.4) (9.5,-1.4)};

\end{tikzpicture}
    \caption{This figure illustrates the operation of the tree decoder. The inner decoder $\mathcal{C}^{-1}$ produces $L$ lists of $K$ messages each. The outer tree decoder then finds parity consistent paths across lists to extract valid messages. }
    \label{fig:slot_decoding}
\end{figure}

Recall that the receiver is tasked with producing an unordered list of all the transmitted messages. 
A naive way to do this is to have the inner and outer decoders operate independently of each other. 
That is, the inner decoder is run on each of the $L$ fragments in $\yv$ to produce $L$ estimates of the outer-encoded codewords. 
Since $\mathcal{C}$ has the property that, given a linear combination of its codewords, the constituent input signals may be recovered with high probability, the aggregate signal in every slot can be expanded into a list of $K$ encoded fragments $\{\hat{\vv}_j(\ell) : j \in [K]\}$. 
It is pertinent to remind the reader that $\hat{\vv}_j(\ell)$ does not necessarily correspond to the message sent by user~$j$ as the receiver has no way of connecting a received message to an active user within URA. 
At this point, the receiver has $L$ lists $\mathcal{L}_1, \mathcal{L}_2, \hdots, \mathcal{L}_{L}$, each with $K$ outer-encoded fragments. 
From these lists, the receiver must estimate the $K$ messages sent by the active devices during the frame.
This is done by running the tree decoder on the $L$ lists to find parity-consistent paths across lists. 
Specifically, the tree decoder first selects a root fragment from list $\mathcal{L}_1$ and computes the corresponding parity section $\pv(2)$. 
The tree decoder then branches out to all fragments in list $\mathcal{L}_2$ whose parity sections match $\pv(2)$; each match creates a parity consistent partial path. 
This process repeats until the last list $\mathcal{L}_{L}$ is processed. 
At this point, if there is a single path from $\mathcal{L}_1$ to $\mathcal{L}_{L}$, the message created by that path is deemed valid and stored for further processing; if there are multiple parity-consistent paths from a given root fragment or no parity consistent paths from a given root fragment, a decoding failure is declared. 
Fig.~\ref{fig:slot_decoding} illustrates this process. 

While intuitive, this strategy is sub-optimal because information is not being shared by the inner and outer decoders. 
If the inner and outer decoders were to operate concurrently, the output of the outer decoder could be used to reduce the search space of the inner decoder, thus guiding the convergence of the inner decoder to a parity consistent solution. 
This would also reduce the search space of the inner code, thus providing an avenue for reducing decoding complexity \cite{amalladinne2020enhanced}, \cite{amalladinne2021mimo}. 
Explicitly, assume that immediately after the inner decoder produces list $\mathcal{L}_\ell$, the outer decoder finds all parity-consistent partial paths from the root node to stage~$\ell$.
Each of these $R$ parity consistent partial paths has an associated parity section $\pv_r(\ell+1)$. 
Furthermore, it is known that only those fragments in $\mathcal{L}_{\ell+1}$ that contain one of the $\{\pv_r(\ell+1) : r \in [R]\}$ admissible parity sections may be part of the $K$ transmitted messages.
Thus, when producing $\mathcal{L}_{\ell+1}$, the search space of the inner decoder may be reduced drastically to just the subset for which fragments contain an admissible parity section $\pv_r(\ell+1)$. 

This algorithmic enhancement has the potential to simultaneously reduce decoding complexity and improve PUPE performance.
Still, a precise characterization of the benefits of this enhanced algorithm depends on the inner code chosen. 
We now consider two situations in which this algorithm may be applied: Coded Compressed Sensing (CCS) \cite{amalladinne2019coded} and CCS for massive MIMO \cite{fengler2019mimo}. 
For each of the considered schemes, the complexity reduction and performance improvements are quantified. 
We emphasize that this algorithmic enhancement is applicable to other scenarios beyond those considered in this paper; one such example is the CHIRRUP scheme presented by Calderbank and Thompson in \cite{calderbank2018chirrup}.

\section{Case Study 1: Coded Compressed Sensing}
\label{section:CCS}

In recent years, CCS has emerged as a practical scheme for URA that offers good performance with low complexity \cite{amalladinne2019coded, amalladinne2020unsourced, ebert2020hybrid, ebert2021stochastic}. 
Though many variants of CCS have emerged, we will focus on the original version published by Amalladinne et al.\ in \cite{amalladinne2019coded}. 
At its core, CCS seeks to exploit a connection between URA and compressed sensing (CS). 
This connection may be understood by transforming a $B$-bit message $\wv$ into a length $2^B$ index vector $\mv$; the single non-zero entry therein is a one at location $[\wv]_2$, which is the binary message $\wv$ interpreted as a radix-$10$ integer.
This bijection is denoted $f(x)$.
The vector $\mv$ may then be compressed into signal $\xv$  using sensing matrix $\Am$ and transmitted over a noisy channel.
The multiple access channel naturally adds the sent signals from the active devices.
At the receiver, the original signals may be recovered from $\yv$ using standard CS recovery techniques such as non-negative least-squares (NNLS) or least absolute shrinkage and selection operator (LASSO). 
However, for messages of even modest lengths, the size of $\xv$ is too large for standard CS solvers to handle.
To circumvent this challenge, a divide and conquer approach can be employed. 

In CCS, the inner code $\mathcal{C}$ consists of the CS encoder and the outer tree code $\mathcal{T}$ is identical to that presented in Section~\ref{section:SystemModel}. 
Note that there is an additional step between $\mathcal{T}$ and $\mathcal{C}$: the outer-encoded message $\vv$ is transformed into the inner code input $\mv$ via the bijection described above. 
Furthermore, $\mathcal{C}$ has the property that, given a linear combination of its codewords, the corresponding set of $K$ one-sparse constituent inputs may be recovered with high probability. 
This, combined with the assumption that $\Pr(\wv_i = \wv_j) < \epsilon$ for $i \neq j$, makes CCS an eligible candidate for the enhanced decoding algorithm described previously. 
We review below the CCS encoding and decoding operations.

\subsection{CCS Encoding}

When user~$j$ wishes to transmit a message to the central base station, it encodes its message in the following manner. 
First, it breaks its $B$-bit message into $L$ fragments and outer-encodes the $L$ fragments using the tree code described in Section~\ref{section:SystemModel}; this yields outer codeword $\vv_j$.
Recall that fragment $\ell$ has $m_\ell$ information bits and $l_\ell$ parity bits.
We emphasize that $m_\ell + l_\ell = v_{\ell}$ is constant for all sections in CCS, but the ratio of $m_\ell$ to $l_\ell$ is subject to change.
Fragment $\vv_j(\ell)$ is then converted into a length $2^{m_\ell + l_\ell}$ index vector, denoted by $\mv_j(\ell)$, and compressed using sensing matrix $\Am$ into vector $\xv_j(\ell)$. 
Within the next transmission frame, user~$j$ transmits its encoded fragments across the GMAC with all other active users.
At the base station, the received vector associated with slot~$\ell$ assumes the form
\begin{equation}
\yv(\ell) = \left( \sum_{j \in [K]} d \Am \mv_j(\ell) \right) + \zv(\ell)
= d \Am \left(\sum_{j \in [K]} \mv_j(\ell) \right) + \zv(\ell)
\end{equation}
where $\zv(\ell)$ is a vector of Gaussian noise with standard normal components and $d$ reflects the transmit power.
This is a canonical form of a $K$-sparse compressed vector embedded in Gaussian noise.

\subsection{CCS Decoding}

CCS decoding begins by running a standard CS solver such as NNLS or LASSO on each section to produce $L$ $K$-sparse vectors. 
The $K$ indices in each of these $L$ slots are converted back to binary representations using $f^{-1}(x)$, and the tree decoder is run on the resultant $L$ lists to produce estimates of the transmitted messages. 

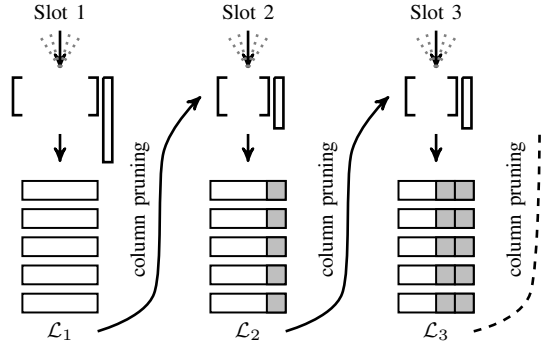
\begin{figure}[htb]
    \centering
    \begin{tikzpicture}
[font=\footnotesize, draw=black, line width=0.75pt,>=stealth',
sub0/.style={rectangle, draw, inner sep=0pt, minimum width=10mm, minimum height=2.5mm},
parity/.style={rectangle, draw, fill=lightgray, inner sep=0pt, minimum size=2.5mm}]

\node (cs1) at (0.00,5.875) {Slot~1};
\node (cs2) at (2.50,5.875) {Slot~2};
\node (cs3) at (5.00,5.875) {Slot~3};

\foreach \v in {0.00,2.50,5.00} {
  \draw[->, line width=1pt]  (\v,4.25) -- (\v,3.875);
  \draw[->, line width=1pt]  (\v,5.625) -- (\v,5.125);
  \draw[dotted, line width=1pt, draw=gray]  (\v-0.25,5.525) -- (\v,5.125);
  \draw[dotted, line width=1pt, draw=gray]  (\v-0.125,5.575) -- (\v,5.125);
  \draw[dotted, line width=1pt, draw=gray]  (\v+0.125,5.575) -- (\v,5.125);
  \draw[dotted, line width=1pt, draw=gray]  (\v+0.25,5.525) -- (\v,5.125);
}

\foreach \v in {0.00} {
  \draw[line width=1pt] (\v-0.5,5) -- (\v-0.625,5) -- (\v-0.625,4.5) -- (\v-0.5,4.5);
  \draw[line width=1pt] (\v+0.375,5) -- (\v+0.5,5) -- (\v+0.5,4.5) -- (\v+0.375,4.5);
  \draw[line width=1pt] (\v+0.575,5) -- (\v+0.575,3.875) -- (\v+0.695,3.875) -- (\v+0.695,5) -- (\v+0.575,5);
}

\foreach \v in {2.50,5.00} {
  \draw[line width=1pt] (\v-0.275,5) -- (\v-0.4,5) -- (\v-0.4,4.5) -- (\v-0.275,4.5);
  \draw[line width=1pt] (\v+0.15,5) -- (\v+0.275,5) -- (\v+0.275,4.5) -- (\v+0.15,4.5);
  \draw[line width=1pt] (\v+0.35,5) -- (\v+0.35,4.325) -- (\v+0.47,4.325) -- (\v+0.47,5) -- (\v+0.35,5);
}

\foreach \p/\c in {3.50/1, 3.125/2, 2.75/3, 2.375/4, 2/5} {
  \node[sub0] (subcs1\c) at (0.0,\p) {};
  \node[sub0] (subcs2\c) at (2.50,\p) {};
  \node[parity] (parity0\c) at (2.875,\p) {};
  \node[sub0] (subcs3\c) at (5.00,\p) {};
  \node[parity] (parity1\c) at (5.125,\p) {};
  \node[parity] (parity2\c) at (5.375,\p) {};
}

\node (list1) at (0.00,1.625) {$\mathcal{L}_{1}$};
\node (list2) at (2.50,1.625) {$\mathcal{L}_{2}$};
\node (list3) at (5.00,1.625) {$\mathcal{L}_{3}$};

\draw [line width=1pt,->] plot[smooth, tension=.5] coordinates {(0.5,1.625) (1.25,2.125) (1.375,4) (1.875,4.75)};
\draw [line width=1pt,->] plot[smooth, tension=.5] coordinates {(3.0,1.625) (3.75,2.125) (3.875,4) (4.375,4.75)};
\draw [line width=1pt,dashed] plot[smooth, tension=.5] coordinates {(5.5,1.625) (6.25,2.125) (6.375,4.25)};

\node[rotate=90] (prune1) at (1.0625,3.25) {column pruning};
\node[rotate=90] (prune2) at (3.5625,3.25) {column pruning};
\node[rotate=90] (prune3) at (6.0625,3.25) {column pruning};
\end{tikzpicture}
    \caption{This figure illustrates the enhanced decoding algorithm applied to CCS. After recovering $\mathcal{L}_{\ell}$, the sensing matrix $\Am$ is pruned so that list $\mathcal{L}_{\ell+1}$ only contains parity-consistent fragments. }
    \label{fig:enhanced_ccs_diagram}
\end{figure}

This process may be improved by applying the proposed enhanced decoding algorithm,
which proceeds as follows for CCS. 
The inner CS solver first recovers section~$1$, and then computes the set of possible parity patterns for section~$2$, denoted by $\mathcal{P}_{2}$. 
The columns of $\Am$ are then pruned dynamically to remove all columns associated with inadmissible parity patterns in section~$2$.
This reduces the number of columns of $\Am$ from $2^{m_1+l_1}$ to $2^{m_1}|\mathcal{P}_1|$ \cite{amalladinne2020enhanced}.
Section~$2$ is then recovered, and the process repeats itself until section $L$ has been decoded; at this point, valid paths through the $L$ lists are identified and the list of estimated transmitted messages is finalized. 
Fig.~\ref{fig:enhanced_ccs_diagram} illustrates this process.

\subsection{Results}

As previously mentioned, the algorithmic enhancement presented in this article has the potential to improve both the performance and the computational complexity of concatenated coding schemes. 
Being URA scheme, the performance of CCS is evaluated with respect to the per-user probability of error (PUPE), which is defined as 
\begin{equation}
    \label{eq:pupe}
    P_e = \frac{1}{K} \sum_{j \in [K]} \Pr \left( \wv_j \notin \hat{W}(\yv) \right)
\end{equation}
where $\hat{W}(\yv)$ is the estimated list of transmitted messages, with at most $K$ items.
Since many different CS solvers with varying computational complexities may be employed within the CCS framework, the complexity reduction offered by the enhanced decoding algorithm will be quantified by counting the number of columns removed from the matrix $\Am$.

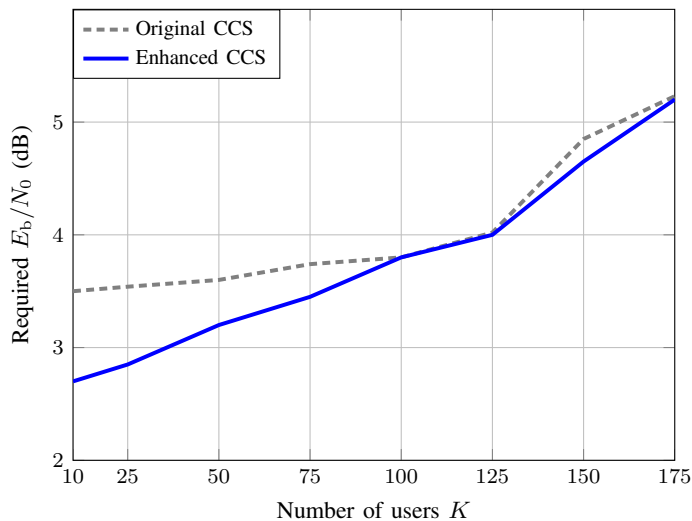
\begin{figure}[htb]
    \centering
    \begin{tikzpicture}

\begin{axis}[%
font=\footnotesize,
width=8cm,
height=6cm,
scale only axis,
xmin=10,
xmax=175,
xtick = {10,25,50,75,...,175},
xlabel={\small Number of users $K$},
xmajorgrids,
ymin=2,
ymax=6,
ytick = {2,...,5},
ylabel={\small Required $E_{\mathrm{b}}/N_0$ (dB)},
ymajorgrids,
legend style={at={(0,1)},anchor=north west,draw=black,fill=white,legend cell align=left}
]


\addplot [color=gray,densely dashed,line width=1.5pt]
  table[row sep=crcr]{
 10 3.5\\ 
25	3.54\\
50	3.6\\
75	3.74\\
100	3.8\\
125	4.02\\
150	4.85\\
175	5.23\\
};
\addlegendentry{Original CCS};

\addplot [color=blue,solid,line width=1.5pt]
  table[row sep=crcr]{
10  2.7\\  
25	2.85\\
50	3.2\\
75	3.45\\
100 3.8\\
125 4\\
150 4.65\\
175 5.2\\
};
\addlegendentry{Enhanced CCS};


\end{axis}

\end{tikzpicture}%
    \caption{This figure shows the required $E_b/N_0$ to obtain a PUPE of $5\%$ vs the number of active users. }
    \label{fig:enhanced_ccs_performance}
\end{figure}

As discussed in \cite{amalladinne2020enhanced}, the column pruning operation has at least four major implications on the performance of CCS. 
These implications are summarized below.
\begin{enumerate}
    \item Many CS solvers rely on iterative methods or convex optimization solvers to recover $\xv$ from $\yv = \Am\xv$. 
    Decreasing the width of $\Am$ will result in a reduction in computational complexity, the exact size of which will depend on the CS solver employed. 
    \item When all message fragments have been correctly recovered for stages $1, 2, \hdots, \ell$, the matrix $\Am$ is pruned in such a way that is perfectly consistent with the true signal.
    In this scenario, the search space for the CS solver is significantly reduced and the performance will improve. 
    \item When an erroneous message fragment has been incorrectly identified as a true message fragment by stage $\ell$, the column pruning operation will guide the CS solver to a list of fragments that is more likely to contain additional erroneous fragments. 
    This further propagates the error and helps erroneous paths stay alive longer. 
    \item When a true fragment is removed from a CS list, its associated parity pattern may be discarded and disappear entirely. 
    This results in the loss of a correct message and additional structured noise which may decrease the PUPE performance of other valid messages. 
\end{enumerate}
Despite having positive and negative effects, the net effect of the enhanced decoding algorithm on the system's PUPE perfomance is positive, as illustrated in Fig.~\ref{fig:enhanced_ccs_performance}. 
This figure was generated by simulating a CCS scenario with $K \in [10:175]$ users, each of which wishes to transmit a $B = 75$ bit message divided into $L = 11$ stages over $22,517$ channel uses. 
NNLS was used as the CS solver. 

From Fig.~\ref{fig:enhanced_ccs_performance}, we gather that the enhanced decoding algorithm reduces the required $E_b/N_0$ by nearly $1$~dB for a low number of users.
Furthermore, for the entire range of number of users considered, the enhanced algorithm is at least as good as the original algorithm and often much better. 

By tracking the expected number of parity-consistent partial paths, it may be possible to compute the expected column reduction ratio at every stage. 
However, this is a daunting task, as explained in \cite{amalladinne2019coded}. 
Instead, we estimate the expected column reduction ratio by applying the analysis from \cite{amalladinne2019coded} with the following simplifying assumptions: 
\begin{itemize}
    \item No two users have the exact same message fragments at any stage: $\wv_i(\ell) \neq \wv_j(\ell)$ whenever $i \neq j$ and for all $\ell \in [L]$.
    \item The inner CS decoder makes no errors in producing lists $\mathcal{L}_1, \hdots, \mathcal{L}_{L}$.
\end{itemize}
Under these assumptions and starting from a designated root node, the number of erroneous paths that survive stage~$\ell$, denoted $L_\ell$, is subject to the following recursion,
\begin{equation} \label{exprec1}
\begin{split}
\mathbb{E} \big[ L_\ell \big]
&= \mathbb{E} [ \mathbb{E} [ L_\ell \mid L_{\ell-1} ] ] \\
&= \mathbb{E} \left[ ( ( L_{\ell-1}+1 ) K-1 ) 2^{-l_{\ell}} \right] \\
&= 2^{-l_{\ell}} K \mathbb{E} [ L_{\ell-1} ] + 2^{-l_{\ell}} (K-1) .
\end{split}
\end{equation}
Using initial condition $\mathbb{E} [L_1] = 0$, we get expected value
\begin{equation}
    \mathbb{E}[L_\ell] = \sum_{q=2}^{\ell} \left(K^{\ell-q}(K-1)\prod_{k=q}^{\ell}2^{-l_{k}} \right) .
\end{equation}
When the matrix $\Am$ is pruned dynamically, then $K$ copies of the tree decoder run in parallel and, as such, the expected number of parity-consistent partial paths at stage~$\ell$ can be expresses as
\begin{equation*}
    P_\ell = K(1 + \mathbb{E}[L_\ell]) .
\end{equation*}

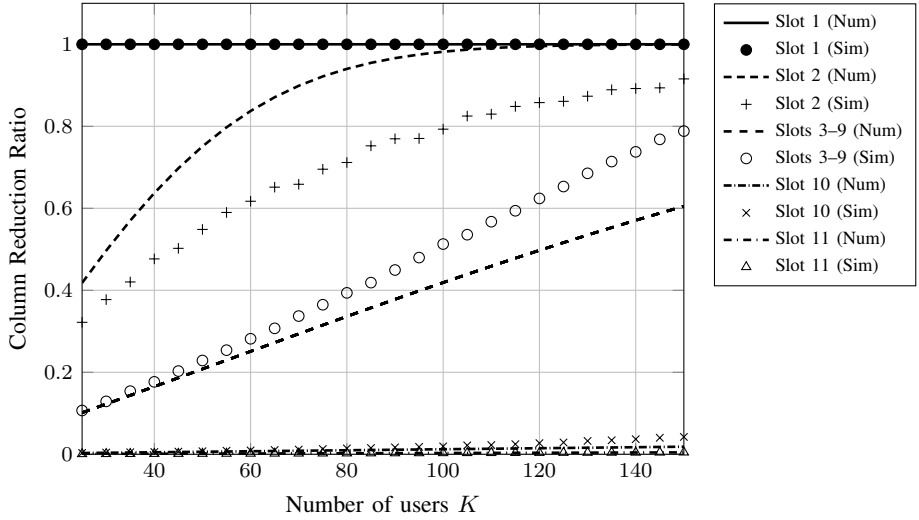
\begin{figure}[htb]
    \centering
    \begin{tikzpicture}

\begin{axis}[%
font=\footnotesize,
width=8cm,
height=6cm,
scale only axis,
xmin=25,
xmax=150,
xlabel={\small Number of users $K$},
xmajorgrids,
ymin=0,
ymax=1.1,
ylabel={\small Column Reduction Ratio},
ymajorgrids,
legend style={at={(1.05,1)},anchor=north west,draw=black, fill=white, legend cell align=left, font=\scriptsize, legend columns=1}
]

\addplot [color=black,solid,line width=1.0pt]
  coordinates {
(25, 1.00000) (30, 1.00000) (35, 1.00000) (40, 1.00000) (45, 1.00000) (50, 1.00000) (55, 1.00000) (60, 1.00000) (65, 1.00000) (70, 1.00000) (75, 1.00000) (80, 1.00000) (85, 1.00000) (90, 1.00000) (95, 1.00000) (100, 1.00000) (105, 1.00000) (110, 1.00000) (115, 1.00000) (120, 1.00000) (125, 1.00000) (130, 1.00000) (135, 1.00000) (140, 1.00000) (145, 1.00000) (150, 1.00000)
};
\addlegendentry{Slot 1 (Num)};

\addplot [only marks,mark=*]
  coordinates {
(25, 1.00000) (30, 1.00000) (35, 1.00000) (40, 1.00000) (45, 1.00000) (50, 1.00000) (55, 1.00000) (60, 1.00000) (65, 1.00000) (70, 1.00000) (75, 1.00000) (80, 1.00000) (85, 1.00000) (90, 1.00000) (95, 1.00000) (100, 1.00000) (105, 1.00000) (110, 1.00000) (115, 1.00000) (120, 1.00000) (125, 1.00000) (130, 1.00000) (135, 1.00000) (140, 1.00000) (145, 1.00000) (150, 1.00000)
};
\addlegendentry{Slot 1 (Sim)};

\addplot [color=black,densely dashed,line width=1.0pt]
  coordinates {
(25, 0.41804) (30, 0.49668) (35, 0.57002) (40, 0.63716) (45, 0.69757) (50, 0.75100) (55, 0.79749) (60, 0.83732) (65, 0.87092) (70, 0.89882) (75, 0.92167) (80, 0.94010) (85, 0.95475) (90, 0.96624) (95, 0.97511) (100, 0.98188) (105, 0.98697) (110, 0.99075) (115, 0.99351) (120, 0.99550) (125, 0.99692) (130, 0.99792) (135, 0.99861) (140, 0.99908) (145, 0.99940) (150, 0.99961)
};
\addlegendentry{Slot 2 (Num)};

\addplot [only marks,mark=+]
  coordinates {
(25,0.3218750) (30,0.3773437) (35,4.203125e-01) (40,4.765625e-01) (45,5.023437e-01) (50,5.484375e-01) (55,5.898438e-01) (60,6.171875e-01) (65,6.515625e-01)
(70,6.585937e-01) (75,6.953125e-01) (80,7.117187e-01) (85,7.523437e-01) (90,7.695313e-01) (95,7.703125e-01)
(100,7.929688e-01) (105,8.250000e-01) (110,8.296875e-01) (115,8.484375e-01) (120,8.578125e-01)(125,8.609e-01)(130,8.734e-01)(135,8.8906e-01)(140,8.9218e-01)(145,8.9375e-01)(150,9.1562e-01)

};
\addlegendentry{Slot 2 (Sim)};

\addplot [color=black,dashed,line width=1.0pt]
  coordinates {
(25, 0.10149) (30, 0.12253) (35, 0.14374) (40, 0.16507) (45, 0.18649) (50, 0.20797) (55, 0.22947) (60, 0.25095) (65, 0.27240) (70, 0.29377) (75, 0.31504) (80, 0.33617) (85, 0.35715) (90, 0.37794) (95, 0.39851) (100, 0.41885) (105, 0.43893) (110, 0.45873) (115, 0.47823) (120, 0.49741) (125, 0.51626) (130, 0.53476) (135, 0.55289) (140, 0.57064) (145, 0.58800) (150, 0.60497)
};
\addlegendentry{Slots 3--9 (Num)};

\addplot [only marks,mark=o]
  coordinates {(25,1.067801e-01) (30,1.290179e-01) (35,1.539342e-01) (40,1.766462e-01) (45,2.029297e-01) (50,2.287109e-01) (55,2.539342e-01) (60,2.817243e-01) (65,3.071429e-01)
(70,3.371094e-01) (75,3.647879e-01) (80,3.937221e-01) (85,4.186663e-01) (90,4.494420e-01) (95,4.797433e-01)
(100,5.127511e-01) (105,5.356027e-01) (110,5.672991e-01) (115,5.939732e-01) (120,6.239955e-01)(125,6.5306e-01)(130,6.8526e-01)(135,7.1395e-01)(140,7.3766e-01)(145,7.6813e-01)(150,7.8825e-01)
};
\addlegendentry{Slots 3--9 (Sim)};

\addplot [color=black,densely dashdotted,line width=1.0pt]
  coordinates {
(25, 0.00306) (30, 0.00367) (35, 0.00428) (40, 0.00489) (45, 0.00551) (50, 0.00612) (55, 0.00674) (60, 0.00735) (65, 0.00797) (70, 0.00858) (75, 0.00920) (80, 0.00981) (85, 0.01043) (90, 0.01104) (95, 0.01166) (100, 0.01228) (105, 0.01290) (110, 0.01352) (115, 0.01413) (120, 0.01475) (125, 0.01537) (130, 0.01599) (135, 0.01661) (140, 0.01723) (145, 0.01785) (150, 0.01847)
};
\addlegendentry{Slot 10 (Num)};

\addplot [only marks,mark=x]
  coordinates {(25,3.332520e-03) (30,4.180908e-03) (35,4.901123e-03) (40,5.511475e-03) (45,6.542969e-03) (50,7.434082e-03) (55,8.605957e-03) (60,9.436035e-03) (65,1.044312e-02)
 (70,1.149292e-02) (75,1.278687e-02) (80,1.430054e-02) (85,1.522217e-02) (90,1.674194e-02) (95,1.811523e-02)
(100,1.921997e-02) (105,2.165527e-02) (110,2.257080e-02) (115,2.467041e-02) (120,2.714844e-02)(125,2.8845e-02)(130,3.2532e-02)(135,3.4167e-02)(140,3.6804e-02)(145,4.0258e-02)(150,4.2333e-02)
};
\addlegendentry{Slot 10 (Sim)};

\addplot [color=black,dashdotted,line width=1.0pt]
  coordinates {
(25, 0.00076) (30, 0.00092) (35, 0.00107) (40, 0.00122) (45, 0.00137) (50, 0.00153) (55, 0.00168) (60, 0.00183) (65, 0.00199) (70, 0.00214) (75, 0.00229) (80, 0.00244) (85, 0.00260) (90, 0.00275) (95, 0.00290) (100, 0.00306) (105, 0.00321) (110, 0.00336) (115, 0.00352) (120, 0.00367) (125, 0.00382) (130, 0.00398) (135, 0.00413) (140, 0.00428) (145, 0.00443) (150, 0.00459)
};
\addlegendentry{Slot 11 (Num)};

\addplot [only marks,mark=triangle]
  coordinates {(25,7.644653e-04) (30,9.170532e-04) (35,1.072693e-03) (40,1.228333e-03) (45,1.380920e-03) (50,1.538086e-03) (55,1.689148e-03) (60,1.841736e-03) (65,1.997375e-03)
 (70,2.151489e-03) (75,2.307129e-03) (80,2.476501e-03) (85,2.627563e-03) (90,2.786255e-03) (95,2.940369e-03)
(100,3.100586e-03) (105,3.288269e-03) (110,3.399658e-03) (115,3.573608e-03) (120,3.771973e-03)(125,3.9337e-03)(130,4.0740e-3)(135,4.2205e-03)(140,4.3854e-03)(145,4.5380e-03)(150,4.7241e-03)
};
\addlegendentry{Slot 11 (Sim)};

\addplot [color=black,dashed,line width=1.0pt]
  coordinates {
(25, 0.10149) (30, 0.12253) (35, 0.14374) (40, 0.16507) (45, 0.18649) (50, 0.20797) (55, 0.22947) (60, 0.25095) (65, 0.27240) (70, 0.29377) (75, 0.31504) (80, 0.33617) (85, 0.35715) (90, 0.37794) (95, 0.39851) (100, 0.41885) (105, 0.43893) (110, 0.45873) (115, 0.47823) (120, 0.49741) (125, 0.51626) (130, 0.53476) (135, 0.55289) (140, 0.57064) (145, 0.58800) (150, 0.60497)
};

\addplot [color=black,dashed,line width=1.0pt]
  coordinates {
(25, 0.10149) (30, 0.12253) (35, 0.14374) (40, 0.16507) (45, 0.18649) (50, 0.20797) (55, 0.22947) (60, 0.25095) (65, 0.27240) (70, 0.29377) (75, 0.31504) (80, 0.33617) (85, 0.35715) (90, 0.37794) (95, 0.39851) (100, 0.41885) (105, 0.43893) (110, 0.45873) (115, 0.47823) (120, 0.49741) (125, 0.51626) (130, 0.53476) (135, 0.55289) (140, 0.57064) (145, 0.58800) (150, 0.60497)
};

\addplot [color=black,dashed,line width=1.0pt]
  coordinates {
(25, 0.10149) (30, 0.12253) (35, 0.14374) (40, 0.16507) (45, 0.18649) (50, 0.20797) (55, 0.22947) (60, 0.25095) (65, 0.27240) (70, 0.29377) (75, 0.31504) (80, 0.33617) (85, 0.35715) (90, 0.37794) (95, 0.39851) (100, 0.41885) (105, 0.43893) (110, 0.45873) (115, 0.47823) (120, 0.49741) (125, 0.51626) (130, 0.53476) (135, 0.55289) (140, 0.57064) (145, 0.58800) (150, 0.60497)
};

\addplot [color=black,dashed,line width=1.0pt]
  coordinates {
(25, 0.10149) (30, 0.12253) (35, 0.14374) (40, 0.16507) (45, 0.18649) (50, 0.20797) (55, 0.22947) (60, 0.25095) (65, 0.27240) (70, 0.29377) (75, 0.31504) (80, 0.33617) (85, 0.35715) (90, 0.37794) (95, 0.39851) (100, 0.41885) (105, 0.43893) (110, 0.45873) (115, 0.47823) (120, 0.49741) (125, 0.51626) (130, 0.53476) (135, 0.55289) (140, 0.57064) (145, 0.58800) (150, 0.60497)
};

\addplot [color=black,dashed,line width=1.0pt]
  coordinates {
(25, 0.10149) (30, 0.12253) (35, 0.14374) (40, 0.16507) (45, 0.18649) (50, 0.20797) (55, 0.22947) (60, 0.25095) (65, 0.27240) (70, 0.29377) (75, 0.31504) (80, 0.33617) (85, 0.35715) (90, 0.37794) (95, 0.39851) (100, 0.41885) (105, 0.43893) (110, 0.45873) (115, 0.47823) (120, 0.49741) (125, 0.51626) (130, 0.53476) (135, 0.55289) (140, 0.57064) (145, 0.58800) (150, 0.60497)
};

\addplot [color=black,dashed,line width=1.0pt]
  coordinates {
(25, 0.10149) (30, 0.12253) (35, 0.14374) (40, 0.16507) (45, 0.18649) (50, 0.20797) (55, 0.22947) (60, 0.25095) (65, 0.27240) (70, 0.29377) (75, 0.31504) (80, 0.33617) (85, 0.35715) (90, 0.37794) (95, 0.39851) (100, 0.41885) (105, 0.43893) (110, 0.45873) (115, 0.47823) (120, 0.49741) (125, 0.51626) (130, 0.53476) (135, 0.55289) (140, 0.57064) (145, 0.58800) (150, 0.60497)
};
\end{axis}

\end{tikzpicture}
    \caption{This figure illustrates the column reduction ratio provided by the enhanced decoding algorithm for each stage of the outer code and a varying number of users. Lines represent numerical results and markers represent simulated results. Clearly, the size of the sensing matrix may be drastically reduced.}
    \label{fig:column_reduction_ratio}
\end{figure}

Under the further simplifying assumptions that all parity patterns are independent and $P_j$ concentrates around its mean, we can approximate the number of admissible parity patterns.
The probability that a particular path maps to a specific parity pattern is $2^{-l_\ell}$ and, hence, the probability that this pattern is not selected by any path become $(1 - 2^{-l_\ell})^{P_\ell}$.
Taking the complement of this event and multiplying by the number of parity patters, we get an approximate expression for the mean number of admissible patterns,
\begin{equation}
|\mathcal{P}_\ell| \approx 2^{l_\ell} \left( 1 - \left( 1 - 2^{-l_\ell} \right)^{P_\ell} \right) .
\end{equation}
Thus, the expected column reduction ratio at slot~$\ell$, denoted $\mathbb{E}[R_\ell]$, is given by (\cite{amalladinne2020enhanced})
\begin{equation}
    \mathbb{E}[R_\ell] = 1 - \left( 1 - 2^{-l_\ell} \right)^{P_\ell}.
\end{equation}
Fig.~\ref{fig:column_reduction_ratio} shows the estimated versus simulated column reduction ratio across stages. 
Overall, the number of columns in $\Am$ can be reduced drastically for some stages, thus significantly lowering the complexity of the decoding algorithm.

\section{Case Study 2: Coded Compressed Sensing for Massive MIMO}
\label{section:CCS_MIMO}

A natural extension of the single-input single-output (SISO) version of CCS proposed in \cite{amalladinne2019coded} is a version of CCS where the base station utilizes $M \gg 1$ receive antennas.
In this scenario, we assume that the receive antennas are sufficiently separated to ensure negligible spatial correlation across channels.
Furthermore, we adopt a block fading model where the channel remains fixed for a coherence period of $n$ channel uses and all coherence blocks are assumed to be completely independent, as in \cite{fengler2019massive}. 
Each active user transmits its message over $L$ coherence blocks, with one coherence block corresponding to each of the $L$ sections described above; thus the total number of channel uses is $N = nL$.
As in SISO CCS, the receiver is tasked with producing an estimated list of the messages transmitted by the collection of active users during a given time instant.
In addition to observing the received signal, the base station has knowledge of the total number of active users, the codes used for encoding messages, and the second-order statistics of MIMO channels.
We note that channel state information (CSI) is not fully known.
Thus the decoding algorithm can be characterized as non-coherent \cite{amalladinne2021mimo}.
The scheme we consider in this work was first presented by Fengler et al.\ in \cite{fengler2019mimo}.

\subsection{MIMO Encoding}

The encoding process for CCS with massive MIMO is analogous to the encoding process for CCS; for a thorough description of this process, please see Section~\ref{section:CCS}. 
However, the signal received by the base station will have a different structure as the base station employs $M$ receive antennas. 
Let $\xv(t, \ell)$ denote the $t$th symbol in block $\ell$ of vector $\xv$. 
Then, the signal observed by the base station is of the form
\begin{equation}
    \yv(t, \ell) = \sum_{j \in [K]} \xv_j(t, \ell)\hv_j(\ell) + \zv(t, \ell) \hspace{5mm} t \in [n], \; \ell \in [L]
\end{equation}
where $\zv(t, \ell)$ is circularly-symmetric complex white Gaussian noise with zero mean and variance $N_0/2$ per dimension and $\hv_j(\ell) \sim \mathcal{CN}(\mathbf{0}, \mathbf{I}_M)$ is a vector of small-scale fading coefficients representing the channel between user~$j$ and the base station's $M$ antennas.

\subsection{MIMO Decoding}

Recall that an URA receiver is tasked with producing an unordered list of the messages transmitted by the collection of active devices. 
To do this, the receiver must first identify the list of fragments transmitted during each of the $L$ coherence blocks and then extract the transmitted messages by finding parity consistent paths across lists. 
The receiver architecture presented in \cite{fengler2019mimo} features a concatenated code, where the inner code $\mathcal{C}$ is decoded using a covariance-based activity detection algorithm and the outer tree code $\mathcal{T}$ is decoded in a manner identical to that presented in Section~\ref{section:SystemModel}. 

\begin{figure}
    \centering
    \begin{tikzpicture}[
  font=\small, >=stealth', line width = 0.75pt,
  wblock/.style={rectangle, minimum height=2mm, draw=black, fill=gray!10},
  pblock/.style={rectangle, minimum height=2mm, draw=black, fill=gray!40}
]

\foreach \y in {0, 0.5, 1, 1.5} {
    \foreach \x in {0, 0.5, 1, 1.5, 2, 2.5, 3, 3.5, 4, 4.5} {
        \draw (\x,\y) to (\x,\y+0.2) to (\x-0.1,\y+0.3) to (\x+0.1,\y+0.3) to (\x,\y+0.2);
     }
}

\draw[|-|] (-0.15, 2.15) -- node[above] {$M$ antennas} (4.65, 2.15);
\draw[->] (-0.35, 1.8) -- node[above,rotate=90] {Time} (-0.35,0);

\end{tikzpicture}
    \caption{This figure illustrates the structure of $\Ym(\ell)$, where the rows correspond to time instants and the columns correspond to receive antennas. }
    \label{fig:mimo_diagram}
\end{figure}
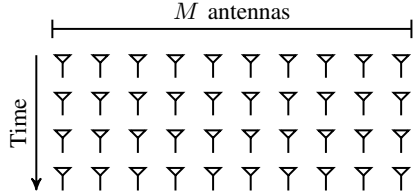

Recall that each active user transforms its outer-encoded message $\vv$ into a $1$-sparse index vector $\mv$.
Let $\{i_j(\ell) : j \in [K]\}$ denote the set of indices chosen by the active users during block $\ell$. 
Then, the signal observed at the base station is of the form
\begin{equation}
    \label{eq:rx_signal}
    \begin{split}
        \Ym(\ell) &= \sum_{j \in [K]} \av_{i_j(\ell)}(\ell)\hv_j(\ell)^\intercal + \Zm(\ell) \\
         &= \Am(\ell)\Gammam(\ell)\Hm(\ell)+\Zm(\ell)
    \end{split}
\end{equation}
where $\Hm(\ell)$ has independent $\mathcal{CN}(0, 1)$ entries, $\Zm$ is independent complex Gaussian noise, and $\Gammam(\ell)$ is a diagonal matrix that indicates which indices have been selected during block $\ell$; that is, $\Gammam(\ell) = \mathrm{diag}(\gamma_0(\ell), \hdots, \gamma_{2^{v_\ell}}(\ell))$ where 
\begin{equation}
    \gamma_i(\ell) = 
    \begin{cases}
    1 & i \in \{i_j(\ell) : j \in [K]\} \\
    0 & \mathrm{otherwise} .
    \end{cases}
\end{equation}
Finally, $\Ym(\ell)$ is a $n \times M$ matrix where the rows of $\Ym(\ell)$ correspond to various time instants and the columns of $\Ym(\ell)$ correspond to the different antennas present at the base station. 
Fig.~\ref{fig:mimo_diagram} illustrates this configuration. 

Determining which fragments were sent during coherence block~$\ell$ is equivalent to estimating $\Gammam(\ell)$. 
This process is referred to as activity detection and may be accomplished through covariance matching when the number of receive antenna is large, as described in \cite{fengler2019mimo}. 
An iterative algorithm for estimating $\Gammam(\ell)$ was first proposed by Fengler in \cite{fengler2019mimo} and is summarized in Algorithm~\ref{alg:activity}.
After the collection of fragments transmitted in each of the $L$ sub-blocks has been recovered by Algorithm~\ref{alg:activity}, tree decoding is employed to disambiguate the collection of transmitted messages. 
\begin{algorithm}[htb]
\caption{Activity Detection via Coordinate Descent}\label{alg:activity}
\begin{algorithmic}[1]
\State \textbf{Inputs}: Sample covariance $\hat{\mathbf{\Sigma}}_{\mathbf{Y}(\ell)} = \frac{1}{M}\mathbf{Y}(\ell)\mathbf{Y}(\ell)^H$
\State \textbf{Initialize}: $\mathbf{\Sigma}_{\ell} = N_0 \mathbf{I}_n$, $\boldsymbol{\gamma}(\ell) = \mathbf{0}$
\For {$i=1,2,\ldots$} 
	\For {$k \in \mathcal{S}_\ell$}
		\State Set $d^* = \frac{\av_k(\ell)^H \mathbf{\Sigma}_\ell^{-1} (\hat{\mathbf{\Sigma}}_{\mathbf{Y}(\ell)}\mathbf{\Sigma}_\ell^{-1} - \mathbf{I}_n)\av_k(\ell)} {(\av_k(\ell)^H \mathbf{\Sigma}_\ell^{-1} \av_k(\ell))^2}$
		\State Update $\gamma_k(\ell) \gets \max \{ \gamma_k(\ell) + d^*, 0 \}$
		\State Update $\mathbf{\Sigma}_\ell^{-1} \gets \mathbf{\Sigma}_\ell^{-1} - \frac{d^*\mathbf{\Sigma}_\ell^{-1}\av_k(\ell)\av_k(\ell)^H\mathbf{\Sigma}_\ell^{-1}}{1 + d^*\av_k(\ell)^H\mathbf{\Sigma}_\ell^{-1}\av_k(\ell)}$
	\EndFor
\EndFor
\State \textbf{Output}: Estimate $\boldsymbol{\gamma}(\ell)$
\end{algorithmic}
\end{algorithm}

As before, it is possible to leverage the enhanced version of the tree decoding process, with its dynamic pruning, to improve performance and lower complexity.
The application of the proposed algorithmic enhancement to the activity detection algorithm may be visualized in the following way. 
Let $\mathcal{S}_\ell$ denote the set of indices to perform coordinate descent over during coherence block $\ell$; in its original formulation, $\mathcal{S}_\ell = [2^{v_\ell}]$. 
After list $\mathcal{L}_1$ has been produced by the activity detection algorithm, the tree decoder can compute the set of all admissible parity patterns $\mathcal{P}_2$ for list $\mathcal{L}_2$; then, $\Am(2)$ may be pruned to only contain those columns corresponding to messages with parity patterns in $\mathcal{P}_2$.
A similar strategy can be applied moving forward, yielding a reduced admissible set $\mathcal{P}_{\ell}$ for parity patterns at stage~$\ell$.
In turn, this reduces the index set $\mathcal{S}_{\ell}$ to 
\begin{equation}
    \mathcal{S}_{\ell} = \{[\wv(\ell)\pv(\ell)]_2 : \wv(\ell) \in \{0, 1\}^{{m_\ell}}, \pv(\ell) \in \mathcal{P}_{\ell} \}
\end{equation}
which may be significantly smaller than $[2^{v_\ell}]$. 
This algorithmic refinement guides the activity detection algorithm to a parity consistent solution and reduces the search space of the inner decoder, thus improving performance significantly \cite{amalladinne2021mimo}.

\subsection{Results}

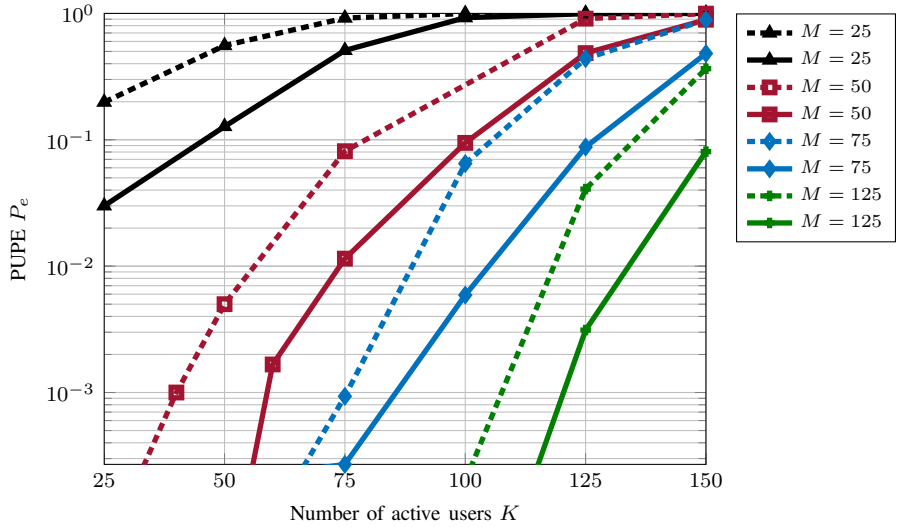
\begin{figure}[htb]
    \centering
    \begin{tikzpicture}

\begin{semilogyaxis}[%
font=\footnotesize,
width=8cm,
height=6cm,
scale only axis,
xmin=25,
xmax=150,
xtick = {25,50,...,150},
xlabel={Number of active users $K$},
xmajorgrids,
ymin=0.00027,
ymax=1,
ytick = {0.0001,0.001,0.01,0.1,1},
ylabel={PUPE $P_e$},
ymajorgrids,
yminorgrids,
legend style={at={(1.05,1)},anchor=north west,draw=black, fill=white, legend cell align=left,font=\scriptsize, legend columns=1}
]

\addplot [color=black,densely dashed,line width=2.0pt,mark=triangle,mark size=1.5pt,mark options={solid}]
  table[row sep=crcr]{
25 0.198400 \\
50 0.5566 \\
75 0.9188 \\
100 0.997 \\
125 0.9992 \\
150 1 \\
};
\addlegendentry{$M=25$};

\addplot [color=black,line width=2.0pt,mark=triangle,mark size=1.5pt,mark options={solid}]
  table[row sep=crcr]{25 0.03 \\
50 0.1274 \\
75 0.5096 \\
100 0.9247 \\
125 0.99416 \\
150 0.999467 \\
};
\addlegendentry{$M=25$};

\addplot [color=mycolor1,densely dashed,line width=2.0pt,mark=square,mark options={solid}]
  table[row sep=crcr]{
25 0\\
32 0.000208\\
40 0.001000\\
50 0.005 \\
75 0.081200 \\
125 0.909120\\
150 0.995333 \\
};
\addlegendentry{$M=50$};

\addplot [color=mycolor1, line width=2.0pt,mark=square,mark options={solid}]
  table[row sep=crcr]{
25 0\\
50 0\\
55 0.000182\\
60 0.001667\\
75 0.011467\\
100 0.094500\\
125 0.485120\\
150 0.891133\\
};
\addlegendentry{$M=50$};

\addplot [color=mycolor2,densely dashed,line width=2.0pt,mark=diamond,mark options={solid}]
  table[row sep=crcr]{
  25 0\\
  50 0\\
  65 0.00021\\
  75 0.000933\\
  100 0.064800\\
  125 0.438560 \\
  150 0.889600\\
};
\addlegendentry{$M=75$};

\addplot [color=mycolor2,line width=2.0pt,mark=diamond,mark options={solid}]
  table[row sep=crcr]{25 0 \\
  50 0.0002 \\
  75 0.00027\\
  100 0.0059 \\
  125 0.0876 \\
  150 0.481667 \\
};
\addlegendentry{$M=75$};

\addplot [color=mycolor3,densely dashed,line width=2.0pt,mark=+,mark options={solid}]
  table[row sep=crcr]{
  100 0.0002 \\
  125 0.040720 \\
  150 0.365333\\
};
\addlegendentry{$M=125$};

\addplot [color=mycolor3,solid,line width=2.0pt,mark=+,mark options={solid}]
  table[row sep=crcr]{100 0 \\
  115 0.000261 \\
  125 0.003120\\
  150 0.080667 \\
};
\addlegendentry{$M=125$};

\end{semilogyaxis}

\end{tikzpicture}%
    \caption{This figure illustrates the performance advantage of applying the enhanced decoding algorithm presented in this paper to CCS for massive MIMO. The dashed line represents the original performance from \cite{fengler2019mimo} and the solid line represents the performance of the enhanced algorithm. }
    \label{fig:ccs_mimo_performance}
\end{figure}

The simulation results presented in this section correspond to a scenario with $K \in [25, 150]$ active users and $M \in [25, 125]$ antennas at the base station. 
Each user encodes their $96$-bit signal into $L = 32$ blocks with $100$ complex channel uses per block. 
The length of the outer-encoded block is $v_\ell = 12$ for all $\ell \in [L]$, and a parity profile of $(l_1, l_2, \hdots, l_{L}) = (0, 9, 9, \hdots, 9, 12, 12, 12)$ is employed.
The energy per bit $E_b/N_0$ is fixed at $0$~dB and the columns of $\Am(\ell)$ are chosen randomly from a sphere of radius $\sqrt{nP}$.
These parameters are chosen to match \cite{fengler2019mimo}. 
Fig.~\ref{fig:ccs_mimo_performance} shows the PUPE of this scheme for a range of active users and several different values of $M$. 
In this figure, the dashed lines represent the performance of the original algorithm and the solid lines represent the performance of the enhanced version with dynamic pruning. 

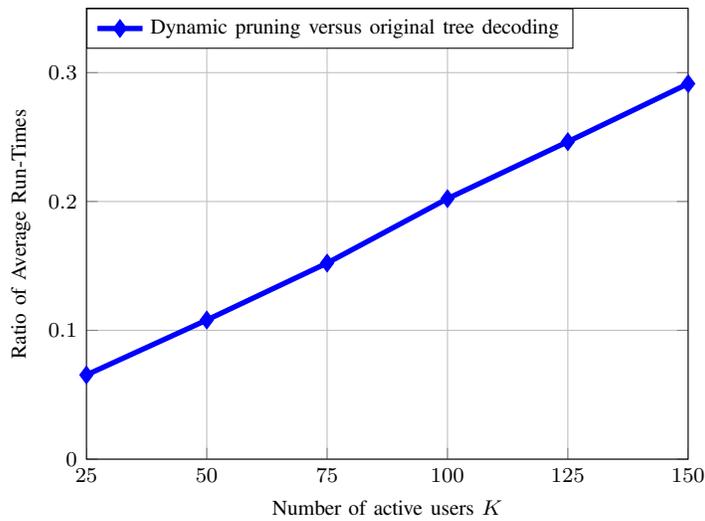
\begin{figure}[htb]
    \centering
    \begin{tikzpicture}

\begin{axis}[%
font=\footnotesize,
width=8cm,
height=6cm,
scale only axis,
xmin=25,
xmax=150,
xtick = {25,50,...,150},
xlabel={Number of active users $K$},
xmajorgrids,
ymin=0,
ymax=0.35,
ytick = {0,0.1,0.2,0.3},
ylabel={Ratio of Average Run-Times},
ymajorgrids,
legend style={at={(0,1)},anchor=north west,draw=black,fill=white,legend cell align=left}
]

\addplot [color=blue,solid,line width=2.0pt,mark=diamond,mark options={solid}]
  table[row sep=crcr]{
  25 0.065355\\
  50 0.1080\\
  75 0.1522\\
  100 0.2022\\
  125 0.2464\\
  150 0.2914\\
};
\addlegendentry{Dynamic pruning versus original tree decoding};






\end{axis}

\end{tikzpicture}%
    \caption{This figure plots the ratio of average runtimes between the enhanced decoding algorithm and the original algorithm. As seen above, dynamic pruning offers a significant reduction in computational complexity compared to standard tree decoding. }
    \label{fig:ccs_mimo_runtimes}
\end{figure}

From Fig.~\ref{fig:ccs_mimo_performance}, we gather that the proposed algorithm reduces the PUPE for a fixed number of active users and a fixed number of antennas at the base station. 
Additionally, this algorithm may be used as a means to reduce the number of antennas required to achieve a target PUPE.
For instance, when $K = 100$, the enhanced algorithm allows for a $23\%$ reduction in the number of antennas at the base station with no degradation in error performance. 
Fig.~\ref{fig:ccs_mimo_runtimes} provides the ratio of average runtimes of the enhanced decoding algorithm versus the original decoding algorithm. 
The enhanced decoding algorithm also offers a significant reduction in computational complexity, especially for a low number of active users.

\section{Conclusion}

In this article, a framework for a concatenated code architecture consisting of a structured inner code and an outer tree code was presented. 
This framework was specifically designed for URA applications, but may find applications in other fields as well. 
An enhanced decoding algorithm was proposed for this framework that promises to improve performance and decrease computational complexity. 
This enhanced decoding algorithm was applied to two URA schemes: coded compressed sensing (CCS) and CCS for massive MIMO. 
In both cases, PUPE performance gains were observed and the decoding complexity was significantly reduced. 

The proposed algorithm is a natural extension of the existing literature. 
From coding theory, we know that there are at least three ways for inner and outer codes to interact. 
Namely, the two codes may operate completely independent of one another in a Forney-style concatenated fashion; this is the style of the original CCS decoder presented in \cite{amalladinne2019coded}. 
Secondly, information messages may be passed between inner and outer decoders as both decoders converge to the correct codeword; this is the style of CCS-AMP which was proposed by Amalladinne et al in \cite{amalladinne2020unsourced}. 
Finally, a successive cancellation decoder may be employed in the spirit of coded decision feedback; this is the style highlighted in this article and considered in \cite{amalladinne2020enhanced, amalladinne2021mimo}. 
Thus, the dynamic pruning introduced in this paper can be framed as an application of coding theoretic ideas to a concatenated coding structure that is common within URA. 

Though the examples presented in this article pertained to CCS, we emphasize that dynamic pruning may be applicable to many algorithms beyond CCS.
For instance, this approach may be relevant to support recovery in exceedingly large dimensions, where a divide and conquer approach is needed.
As long as the inner and outer codes subscribe to the structure described in Section~\ref{section:SystemModel}, this algorithmic enhancement can be leveraged to obtain performance and/or complexity improvements.

\bibliographystyle{IEEEbib}
\bibliography{sensors}

\begin{thebibliography}{10}

\bibitem{polyanskiy2017perspective}
Y.~Polyanskiy,
\newblock ``A perspective on massive random-access,''
\newblock in {\em 2017 IEEE International Symposium on Information Theory
  (ISIT)}, 2017, pp. 2523--2527.

\bibitem{ordentlich2017isit}
O.~Ordentlich and Y.~Polyanskiy,
\newblock ``Low complexity schemes for the random access gaussian channel,''
\newblock in {\em International Symposium on Information Theory (ISIT)}, 2017,
  pp. 2528--2532.

\bibitem{vem2017user}
A.~Vem, K.~R. Narayanan, J.~Cheng, and J.-F. Chamberland,
\newblock ``A user-independent serial interference cancellation based coding
  scheme for the unsourced random access gaussian channel,''
\newblock in {\em Information Theory Workshop (ITW)}. IEEE, 2017, pp. 121--125.

\bibitem{pradhan2020sparse}
A.~Pradhan, V.~Amalladinne, A.~Vem, K.~R. Narayanan, and J.-F. Chamberland,
\newblock ``Sparse {IDMA}: A joint graph-based coding scheme for unsourced
  random access,'' 2020.

\bibitem{marshakov2019polar}
E.~Marshakov, G.~Balitskiy, K.~Andreev, and A.~Frolov,
\newblock ``A polar code based unsourced random access for the gaussian mac,''
\newblock in {\em 2019 IEEE 90th Vehicular Technology Conference
  (VTC2019-Fall)}, 2019, pp. 1--5.

\bibitem{pradhan2019polar}
A.~K. Pradhan, V.~K. Amalladinne, K.~R. Narayanan, and J.-F. Chamberland,
\newblock ``Polar coding and random spreading for unsourced multiple access,''
\newblock in {\em ICC 2020 - 2020 IEEE International Conference on
  Communications (ICC)}, 2020, pp. 1--6.

\bibitem{pradhan2019joint}
A.~Pradhan, V.~Amalladinne, A.~Vem, K.~R. Narayanan, and J.-F. Chamberland,
\newblock ``A joint graph based coding scheme for the unsourced random access
  gaussian channel,''
\newblock in {\em 2019 IEEE Global Communications Conference (GLOBECOM)}, 2019,
  pp. 1--6.

\bibitem{pradhan2021ldpc}
A.~K. Pradhan, V.~K. Amalladinne, K.~R. Narayanan, and J.-F. Chamberland,
\newblock ``Ldpc codes with soft interference cancellation for uncoordinated
  unsourced multiple access,''
\newblock in {\em ICC 2021 - IEEE International Conference on Communications},
  2021, pp. 1--6.

\bibitem{amalladinne2019coded}
Vamsi~K. Amalladinne, Jean-Francois Chamberland, and Krishna~R. Narayanan,
\newblock ``A coded compressed sensing scheme for unsourced multiple access,''
\newblock {\em IEEE Transactions on Information Theory}, vol. 66, no. 10, pp.
  6509--6533, 2020.

\bibitem{fengler2019sparcs}
A.~Fengler, P.~Jung, and G.~Caire,
\newblock ``{SPARC}s for unsourced random access,''
\newblock {\em arXiv preprint arXiv:1901.06234}, 2019.

\bibitem{amalladinne2020unsourced}
V.~K. Amalladinne, A.~K. Pradhan, C.~Rush, J.-F. Chamberland, and K.~R.
  Narayanan,
\newblock ``Unsourced random access with coded compressed sensing: Integrating
  amp and belief propagation,'' 2020.

\bibitem{calderbank2018chirrup}
R.~Calderbank and A.~Thompson,
\newblock ``{CHIRRUP}: {A} practical algorithm for unsourced multiple access,''
\newblock {\em Information and Inference}, December 2019,
\newblock iaz029.

\bibitem{ebert2020hybrid}
J.~R. Ebert, V.~K. Amalladinne, J.-F. Chamberland, and K.~R. Narayanan,
\newblock ``A hybrid approach to coded compressed sensing where coupling takes
  place via the outer code,''
\newblock in {\em ICASSP 2021 - 2021 IEEE International Conference on
  Acoustics, Speech and Signal Processing (ICASSP)}, 2021, pp. 4770--4774.

\bibitem{ebert2021stochastic}
J.~R. Ebert, V.~K. Amalladinne, S.~Rini, J.-F. Chamberland, and K.~R.
  Narayanan,
\newblock ``Stochastic binning and coded demixing for unsourced random
  access,''
\newblock in {\em 2021 IEEE 22nd International Workshop on Signal Processing
  Advances in Wireless Communications (SPAWC)}, 2021, pp. 351--355.

\bibitem{decurninge2020tensorbased}
A.~Decurninge, I.~Land, and M.~Guillaud,
\newblock ``Tensor-based modulation for unsourced massive random access,''
\newblock {\em IEEE Wireless Communications Letters}, vol. 10, no. 3, pp.
  552--556, 2021.

\bibitem{shyianov2020massive}
Volodymyr Shyianov, Faouzi Bellili, Amine Mezghani, and Ekram Hossain,
\newblock ``Massive unsourced random access based on uncoupled compressive
  sensing: Another blessing of massive mimo,''
\newblock {\em IEEE Journal on Selected Areas in Communications}, vol. 39, no.
  3, pp. 820--834, 2021.

\bibitem{han2021sparsekron}
Z.~Han, X.~Yuan, C.~Xu, S.~Jiang, and X.~Wang,
\newblock ``Sparse kronecker-product coding for unsourced multiple access,''
\newblock {\em IEEE Wireless Communications Letters}, vol. 10, no. 10, pp.
  2274--2278, 2021.

\bibitem{li2020sparcldpc}
T.~Li, Y.~Wu, M.~Zheng, D.~Wang, and W.~Zhang,
\newblock ``Sparc-ldpc coding for mimo massive unsourced random access,''
\newblock in {\em 2020 IEEE Globecom Workshops (GC Wkshps}, 2020, pp. 1--6.

\bibitem{xie2020correlatedmimo}
X.~Xie, Y.~Wu, J.~Gao, and W.~Zhang,
\newblock ``Massive unsourced random access for massive mimo correlated
  channels,''
\newblock in {\em GLOBECOM 2020 - 2020 IEEE Global Communications Conference},
  2020, pp. 1--6.

\bibitem{liang2021iterativemimo}
Z.~Liang and J.~Zheng,
\newblock ``Iterative receiver of uplink massive mimo unsourced random
  access,''
\newblock in {\em 2021 International Wireless Communications and Mobile
  Computing (IWCMC)}, 2021, pp. 122--127.

\bibitem{fengler2019massive}
Alexander Fengler, Giuseppe Caire, Peter Jung, and Saeid Haghighatshoar,
\newblock ``Massive mimo unsourced random access,'' 2019.

\bibitem{fengler2019mimo}
Alexander Fengler, Saeid Haghighatshoar, Peter Jung, and Giuseppe Caire,
\newblock ``Non-bayesian activity detection, large-scale fading coefficient
  estimation, and unsourced random access with a massive mimo receiver,''
\newblock {\em IEEE Transactions on Information Theory}, vol. 67, no. 5, pp.
  2925–2951, May 2021.

\bibitem{truhachev2021lowcomplexura}
D.~Truhachev, M.~Bashir, A.~Karami, and E.~Nassaji,
\newblock ``Low-complexity coding and spreading for unsourced random access,''
\newblock {\em IEEE Communications Letters}, vol. 25, no. 3, pp. 774--778,
  2021.

\bibitem{amalladinne2020enhanced}
V.~K. Amalladinne, J.-F. Chamberland, and K.~R. Narayanan,
\newblock ``An enhanced decoding algorithm for coded compressed sensing,''
\newblock in {\em International Conference on Acoustics, Speech and Signal
  Processing (ICASSP)}. IEEE, 2020.

\bibitem{amalladinne2021mimo}
Vamsi~K. Amalladinne, Jean-Francois Chamberland, and Krishna~R. Narayanan,
\newblock ``Coded compressed sensing with successive cancellation list decoding
  for unsourced random access with massive mimo,'' 2021.

\end{thebibliography}

\end{document}